# Fast and slow nonlinearities in ENZ materials

*Jacob B. Khurgin*, Matteo Clerici, Nathaniel Kinsey*

Jacob B Khurgin
Department of Electrical and Computer Engineering, Johns Hopkins University, Baltimore, Maryland 21218, USA  E-mail jakek@jhu.edu

Matteo Clerici,
School of Engineering, University of Glasgow , G128QQ, UK.

Nathaniel Kinsey,
Department of Electrical and Computer Engineering, Virginia Commonwealth University, Richmond, Virginia 23284, USA

## Abstract

Novel materials, with enhanced light-matter interaction capabilities, play an essential role in achieving the lofty goals of nonlinear optics. Recently, Epsilon-Near-Zero (ENZ) media have emerged as a promising candidate to enable the enhancement of several nonlinear processes including refractive index modulation and harmonic generation. Here, we analyze the optical nonlinearity of ENZ media to clarify the commonalities with other nonlinear media and its unique properties. We focus on transparent conducting oxides (TCOs) as the family of ENZ media with near zero permittivity in the near-infrared (telecom) band. We investigate the instantaneous and delayed nonlinearities. By identifying their common origin from the band nonparabolicity, we show that their relative strength is entirely determined by a ratio of the energy and momentum relaxation (or dephasing) times. Using this framework, we compare ENZ materials against the many promising nonlinear media that have been investigated in literature and show that while ENZ materials do not radically outpace the strength of traditional materials in either the fast or slow nonlinearity, they pack key advantages such as an ideal response time, intrinsic slow light enhancement, and broadband nature in a compact platform making them a valuable tool for ultrafast photonics applications for decades to come.

1. **Introduction**

Nonlinear optics has been and remains a fascinating field since the 1960s, ignited shortly after the invention of the laser [1, 2]. In the most general form, nonlinear optics enables the optical control of light in the temporal and frequency domains, resulting in phenomena such as the intensity-dependent index, nonlinear and multiphoton absorption, self and cross-phase modulation, wave mixing, sum and difference-frequency generation, and multiple harmonic generation. Over the last 60 years, advances made in the study and application of these effects have culminated in notable practical solutions including Kerr mode-locking [3], optical frequency combs [4], optical parametric generators and oscillators [5], high-harmonic generation [6], and others.

Yet from the inception of the field, it has been understood that these nonlinear photon-photon interactions are mediated by materials. As such, the development of new and high-performance nonlinear optical media has been intertwined with the discovery of new processes and the realization of practical solutions. Despite decades of study and the several potentially transformative developments – think of organics [7], semiconductor quantum wells and superlattices [8], plasmonic metals [9], carbon nanotubes [10], and 2D materials [11, 12] – the list of materials playing a leading role in nonlinear optical applications has not significantly expanded beyond what was available in the 1980s. Indeed, second-order processes are mostly enabled by conventional crystals like $LiNbO_3$ [13], BBO [14], and KTP [15] from UV to near-IR, and $AgGaSe_2$ and other chalcogenides for mid- and far-IR [16]. Similarly, applications relying on third-order processes are mostly based on silica fiber [17], SiN [18], Si [19], III-V [20] and II-VI [21] semiconductors as well as chalcogenide glasses [22]. This illustrates the extreme difficulty to find a material that can satisfy the sometimes opposing requirements imposed by applications, such as large nonlinearity, speed, wide optical bandwidth, high damage threshold, thermal stability, and others.

In the last few years, a new class of promising nonlinear materials has emerged, characterized by a refractive index (and hence the real part of dielectric constant) that approaches zero, called "*Epsilon-Near-Zero*" or ENZ materials. The real part of ε may approach zero in many systems, such as in polar crystals near the longitudinal optical phonon (SiC) and in metals at the plasma frequency. In the visible and near IR range ENZ materials are predominantly realized with heavily (degenerately) doped semiconductors such as indium tin oxide (ITO), aluminum zinc oxide (AZO), gallium doped zinc oxide (GZO), and others in the transparent conducting oxide (TCO) family [23]. We shall focus on TCOs from here on, due to their relevance to ultrafast photonics applications in the visible and telecom spectral ranges.

Simply, the concept of ENZ can be understood by considering the contributions to the dielectric constant of the TCO described by the well-known Drude-Lorentz relative permittivity $\varepsilon(\omega) = \varepsilon_\infty - Ne^2 / \varepsilon_0 m^* \omega(\omega + i\gamma)$ where $m^*$ is effective mass. When the free carrier density is sufficiently large, the negative contribution due to free electrons, and the positive contribution $\varepsilon_\infty$, arising from bound electrons, nearly cancel each other at a particular wavelength and a number of interesting effects ensue such as low group velocity, enhanced diffraction, and in particular, enhanced nonlinear phenomena. In the case of TCOs, when the carrier density reaches $10^{20}$ - $10^{21}$ $cm^{-3}$, the cancellation occurs in the telecommunications spectral range, which has clear practical implications. As a result, ENZ materials have experimentally demonstrated enhancement to several nonlinear phenomena including (but not limited to) third [24] and higher [25] harmonic generation and cross-phase modulation [26, 27]

In experiments dealing with switching [28], phase conjugation [29], negative refraction [30], cross-phase modulation [31], and adiabatic frequency shifts [32, 33], the origin of the nonlinearity is generally explained by an increase of the electron temperature that exhibits a temporal response of a few hundred femtoseconds, commensurate with the rate of thermal relaxation of hot carriers – a "*slow*" nonlinearity (although it is still fast, on the order of 1 ps, in an absolute sense). At the same time, experiments demonstrating enhanced harmonic generation [24, 25, 34] cannot be explained by such a slow nonlinearity and require a "*fast*" or instantaneous temporal effect.

It appears then that ENZ materials possess more than one kind of nonlinearity, and it is not clear if and how they are related to each other. Here we seek to solve this conundrum, discussing the relation between fast and slow nonlinearities in TCOs. In addition, we utilize this framework to compare the performance of ENZ materials to the wide range of previous nonlinear materials. To do so, in Section 2, we outline the common origin of "*fast*" and "*slow*" optical nonlinearities in dielectrics and establish how their strengths are related. In Section 3 we derive equations for carrier transport in a nonparabolic band. Based on that, in Section 4 we derive the expression and estimate the magnitude of "*fast*" nonlinearity in ENZ and show that it is of the same order of magnitude as fast nonlinearity in conventional materials at the same wavelength. In Section 5 we estimate the "*slow*" nonlinearity in ENZ material, and show that it has the same physical origin as the fast nonlinearity and that the strengths of the two components are related in exactly the same way as in more conventional nonlinear materials. In Section 6 the comparison between different nonlinearities is performed using common figures of merit and the niches where ENZ materials have advantages are identified. Finally, the conclusions are drawn in Section 7.

2. **Fast and slow nonlinearities- general discussion**

Third order nonlinearities are the lowest order nonlinear phenomena intrinsic to all media (with or without inversion symmetry). Any third-order optical process is characterized by the nonlinear susceptibility $\chi^{(3)}(\omega_4 = \omega_1 \pm \omega_2 \pm \omega_3)$ which reveals that in essence it is a four-wave interaction. However, since two or more of the waves can be degenerate, this susceptibility can describe self- and cross-phase modulation, third harmonic generation, and quite a few other nonlinear phenomena. Furthermore, being a complex number, the susceptibility can describe optically induced change in both refraction and absorption.

Third-order processes are usually sub-divided into two broad categories – "*fast*" or instantaneous nonlinearities, and "*slow*" or delayed nonlinearities. The *fast* or *instantaneous* nonlinearity is associated with '*virtual*' processes, i.e., not mediated by real transitions. Familiar examples of such nonlinearities are the optical Kerr effect [35] as well as harmonic generation in dielectrics [36] or semiconductors excited well below the bandgap [35]. These processes are usually associated with low optical losses and a limited magnitude of the nonlinearity. The second is the variety of *slow* nonlinearities where real excitations take place and persist over a certain lifetime, which may be the recombination time or thermal diffusion time. A typical example is the nonlinear index for semiconductors excited near the bandgap [37]. The magnitude of these nonlinearities is relatively high and they exhibit resonances. For these processes the response is slow, on the scale of recombination time (picosecond to nanosecond), and losses are large as the nonlinear effects are based on the absorption and excitation of real carriers in the band. The speed can be enhanced by reducing the recombination time, e.g. including defects [38], although this comes, at the expense of a decreased magnitude. Yet, the ability to engineer the response of slow nonlinearity, e.g. by annealing [38, 39], makes them attractive for optical control applications.

While the fast and slow nonlinearities are usually treated differently, they are in fact closely related, and as we show in this section, practically every nonlinear process has both a fast and a slow component.

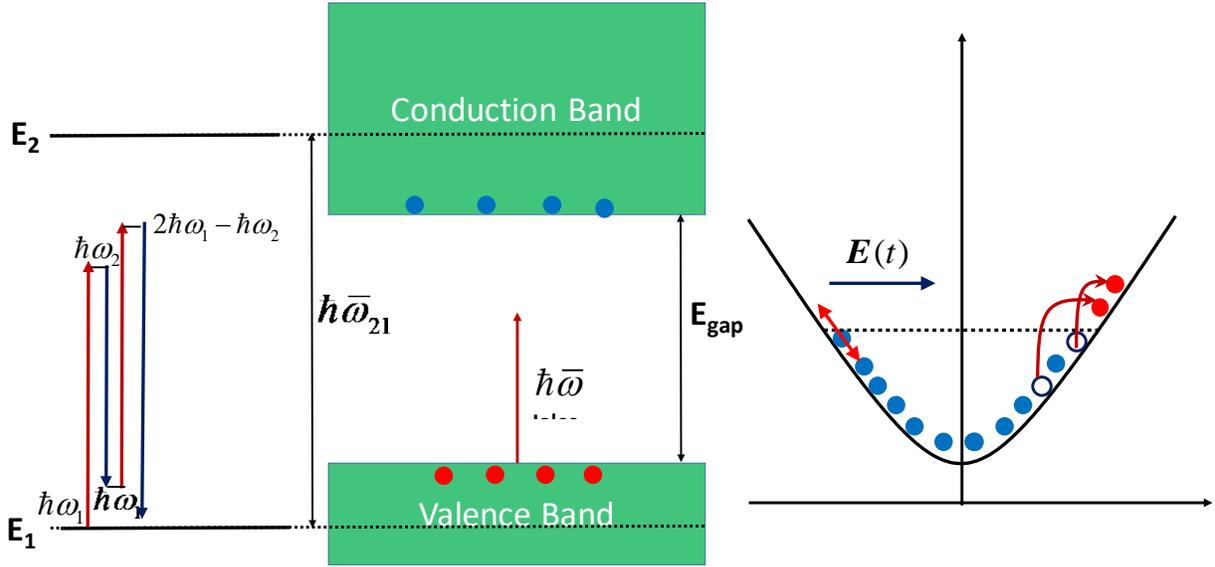

**Figure 1.** (a) two level system and a diagram of a degenerate four wave mixing process in it. (b) two-band solid state nonlinear material that is approximated by a two level system (c) intraband nonlinearity in a nonparabolic conduction band. On the left, electrons near Fermi level move in the nonparabolic energy band in the presence of optical field $E(t)$ – origin of fast nonlinearity. On the right, the electrons get promoted from below the Fermi level to above Fermi level where they are considered "hot" and have higher effective mass-the origin of slow nonlinearity.

The nonlinearity that is perhaps the most familiar originates from the transitions between discrete atomic or molecular energy levels, or between the energy bands in dielectrics or semiconductors. As an example of this process, we consider a degenerate four wave mixing (FWM) in the two level system (**Figure1** a), the process in which two waves with frequencies $\omega_1$ and $\omega_2$ mix and build up a time-dependent polarization oscillating with frequency components $2\omega_2 - \omega_1$ and $2\omega_1 - \omega_2$.

This choice allows us to explore the temporal response of the nonlinear susceptibility $\chi^{(3)}(\omega_3 = \omega_1 - \omega_2 + \omega_1)$ by varying the beat frequency $\Delta\omega = \omega_2 - \omega_1$.

The nonlinear medium can be modeled with a two-level system characterized by the usual set of density matrix equations [40, 41]:

$$\frac{d\Delta\rho}{dt} = -(\Delta\rho - 1)/T_1 + i\frac{\mu_{21} \cdot E}{\hbar}(\rho_{21} - \rho_{12})$$
$$\frac{d\rho_{21}}{dt} = -\gamma\rho_{21}(r) - i\omega_{21}\rho_{21}(r) + i\frac{\mu_{21} \cdot E}{\hbar}\Delta\rho \quad (1)$$
$$\rho_{21} = \rho_{12}^*$$

where $\hbar\omega_{21}$ is the transition energy, $\Delta\rho = \rho_{11} - \rho_{22}$ is the population difference between the lower and upper levels. The off-diagonal terms $\rho_{21}$ and $\rho_{12}$ are sometimes called coherences and describe the net atomic polarization; $\mu_{21} = ez_{21}$ is the transition dipole matrix element; $T_1$ is the inter-level relaxation (recombination) time; $\gamma = 1/T_2$ is the scattering rate, where $T_2$ is the dephasing time. We can write the electric field $E(t) = E_1 e^{-i\omega_1 t} + E_2 e^{-i\omega_2 t} + c.c.$, and in the rotating wave approximation we can expand the matrix elements adopting a perturbative approach:

$$\rho_{21} = \sigma_1 e^{-i\omega_1 t} + \sigma_2 e^{-i\omega_2 t} + \sigma_{112} e^{-i(2\omega_1 - \omega_2)t} + \sigma_{221} e^{-i(2\omega_2 - \omega_1)t} + ...$$
$$\Delta\rho = 1 - \delta_{12} e^{-i(\omega_1 - \omega_2)t} - \delta_{12}^* e^{+i(\omega_1 - \omega_2)t} + ... \quad (2)$$

The first equation illustrates that the net polarization is a sum of both linear (two first terms) and nonlinear polarizations. Here, we have kept the linear polarization terms $\sigma_{1(2)}$ oscillating at $\omega_1, \omega_2$ and FWM terms $\sigma_{112(221)}$ oscillating at $2\omega_2 - \omega_1$ and $2\omega_1 - \omega_2$, which are the subject of this section. However, one should keep in mind that other third-order terms corresponding to third harmonic, sum frequencies, and others are also present. Similarly, the net population difference between states 1 and 2, $\Delta\rho$, can also be expanded into terms which oscillate at different frequencies. Here, we have kept only the second order terms $\delta_{12}$ describing population variation at the beat frequency $\Delta\omega = \omega_1 - \omega_2$ because it is these population pulsations that engender FWM.

Substituting the expansion (2) into (1) and equating the terms oscillating at the same frequency on the l.h.s. and r.h.s. results first in the steady state solution for the linear atomic polarization:

$$\sigma_{1(2)} = \frac{\mu_{21} E_{1(2)} \hbar^{-1}}{(\omega_{21} - \omega_1) - i\gamma}. \tag{3}$$

which is the traditional Lorentzian. Second, the expression for the oscillations of population difference that occur at the beat frequency are found as

$$\delta_{12} = \frac{\omega_1 - \omega_2 + 2i\gamma}{\omega_1 - \omega_2 + i/T_1} \frac{\mu_{21}^2 E_1 E_2 / \hbar^2}{(\omega_{21} - \omega_1 - i\gamma)(\omega_{21} - \omega_2 + i\gamma)}. \tag{4}$$

Once this population beating is mixed with a third electric field at frequency $\omega_1$ or $\omega_2$ the third-order coherence at the intermodulation frequency $2\omega_1 - \omega_2$ or $2\omega_2 - \omega_1$ is produced:

$$\sigma_{112} = \frac{\omega_1 - \omega_2 + 2i\gamma}{\omega_1 - \omega_2 + i/T_1} \frac{\mu_{21}^3 E_1^2 E_2 / \hbar^3}{(\omega_{21} - \omega_1 - i\gamma)(\omega_{21} - \omega_2 + i\gamma)(\omega_{21} - 2\omega_1 + \omega_2 - i\gamma)} \tag{5}$$

Now, the material polarization is $P = N\mu_{21}\rho_{21} + c.c.$, where N is the density of the two-level entities. Therefore, the first-order (linear) polarization can be found as:

$$P^{(1)}(\omega_1) = N\mu_{21}\sigma_1 e^{-\omega_1 t} + c.c. \equiv \varepsilon_0 \chi^{(1)}(\omega_1) E_1^1 e^{-\omega_1 t} + c.c. \tag{6}$$

and the third-order nonlinear polarization responsible for FWM can be found as:

$$P^{(3)}(2\omega_1 - \omega_2) = N\mu_{21}\sigma_{112} e^{-i(2\omega_1 - \omega_2)t} + c.c. \equiv \varepsilon_0 \chi^{(3)}(\omega_3 = \omega_1 - \omega_2 + \omega_1) E_1^2 E_2^* e^{-i(2\omega_1 - \omega_2)t} + c.c. \tag{7}$$

where we have introduced the first- and third-order susceptibilities, $\chi^{(1)}$ and $\chi^{(3)}$, respectively. If we assume that that the frequencies are reasonably close, i.e. $\Delta\omega \ll \omega_1, \omega_2$ then one can approximate all the optical frequencies in the denominator of (5) by the mean frequency $\bar{\omega} = (\omega_1 + \omega_2)/2$, and obtain the expression for third order susceptibility for the FWM process:

$$\chi^{(3)}(\omega_3 = \omega_1 - \omega_2 + \omega_1) \sim \frac{\Delta\omega + 2j\gamma}{\Delta\omega + i/T_1} \frac{N\mu_{21}^4/\varepsilon_0\hbar^3}{\left[(\omega_{21} - \bar{\omega})^2 + \gamma^2\right]^{3/2}} e^{j\Phi} =$$
$$= \left(1 + \frac{2T_1/T_2 - 1}{1 - i\Delta\omega T_1}\right) \frac{N\mu_{21}^4/\varepsilon_0\hbar^3}{\left[(\omega_{21} - \bar{\omega})^2 + \gamma^2\right]^{3/2}} e^{j\Phi} = \chi_{fast}^{(3)} + \chi_{slow}^{(3)}$$

(8)

where $\Phi$ is the phase that is not important for the present discussion.

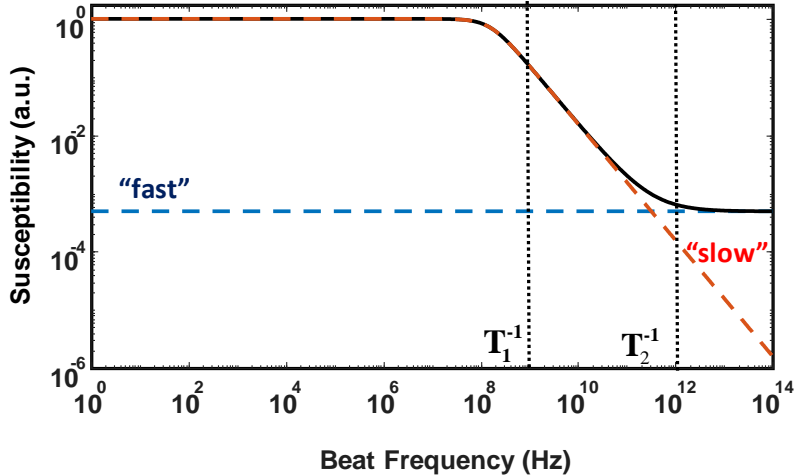

**Figure 2.** Frequency dependence of the FWM susceptibility (solid line) and its fast and slow components (dashed lines) for the material system with $T_1 = 1\ ns$ and $T_2 = 1\ ps$

The nonlinear polarization contains two terms that have common dependence on material parameters but differ in their dependence of beat frequency as shown in **Figure 2**. There is a fast (almost instantaneous) nonlinear response $\chi_{fast}^{(3)}$ that is frequency independent, and a slow response $\chi^{(3)}_{slow}$ whose frequency response has 3dB bandwidth $1/2\pi T_1$. The slow term is related to absorption and the so-called "real" excitations, where electrons get excited (instantaneously) to the upper state and then decay after the time $T_1$. The fast term is is often related to "virtual" excitations, where carriers spend very short time in the upper state, determined by the uncertainty principle.

At low beat frequencies, $\Delta\omega < 1/T_1$, the slow term is larger than the fast one by $2T_1/T_2 - 1$. For most nonlinear media the ratio $2T_1/T_2 \gg 1$ and the slow nonlinearity is much larger than the fast one, with the possible exception of low-pressure atomic gases. However, once the beat frequency $\Delta\omega$ exceeds $2/T_2$, the fast nonlinearity becomes dominant.

Notice that even when the two frequencies are the same and one talks about for example self-phase modulation, there are still two components to $\chi^{(3)}$ – the fast and slow one. It can also been seen that for other nonlinear processes, for example third-harmonic generation, $\Delta\omega$ gets replaced by $2\omega$ in (8) and the "slow" response becomes extremely small, leaving only the instant nonlinearity.

We may also consider the relative magnitudes of the individual contributions. For the "*fast*" effect, the third-order susceptibility can be expressed via the first order susceptibility $\chi^{(1)}(\omega) = \varepsilon(\omega) - 1$ (6) as:

$$\chi^{(3)}_{fast} \sim \chi^{(1)} \frac{e^2 z_{21}^2}{\left[(\omega_{21} - \bar{\omega})^2 + \gamma^2\right]\hbar^2} \qquad (9)$$

where $z_{21}$ is on the scale of the interatomic distance, i.e. 0.1 nm and for the case of condensed matter and $\hbar(\omega_{21} - \omega)$ is on the scale of an eV, while $\chi^{(1)}$ is on the scale of 10 eV which means the scale $\chi^{(3)}_{fast} \sim 10^{19} m^2/V^2$, which is easy to interpret as the inverse square of the internal field in a typical polarizable bond.

Although Equation (8) has been derived in the case of a two-level system, the same model can be applied to describe the nonlinear effects of standard dielectric materials. For these materials the upper and lower states are distributed over the valence and conduction bands that extend over ranges comparable to or even wider than the optical bandgap (width of transmission region). In this case, for transitions that are sufficiently detuned from the absorption edge, $\hbar\omega < E_{gap}$, one can approximate the band by a two-level system with some average optical transition energy, often referred to as the "Penn gap" $\hbar\bar{\omega}_{21}$ [42] which is significantly larger than $E_{gap}$ (by a few eV), see Figure 1b. When $\bar{\omega}_{21}$ is substituted for $\omega_{21}$ in (9), and, since $\hbar\bar{\omega} \leq E_{gap}$, the denominator in (9) is always on the scale of a few eV$^2$ and is nearly frequency independent. This approximation has been used with a great degree of accuracy to describe both the linear [43, 44] and nonlinear [45, 46] properties of many crystals and can be used in conjunction with equation (8) and equation (9) to estimate the fast nonlinearity for dense materials.

For the "slow" nonlinearity, the situation is different because the absorption underpinning the slow effect does not follow the Lorentz behavior seen in the two level system. Rather, the absorption changes with frequency, for example, it decreases exponentially below the bandgap as described by the Urbach rule [47, 48]. Nevertheless, equation (8) can still be used by allowing the effective scattering rate $\gamma = T_2^{-1}$ in (8) to be a function of optical frequency $\gamma(\omega)$ (as was done in [49, 50] where it was shown that the effective scattering rate decreases almost exponentially with detuning from the bandgap $E_{gap} - \hbar\omega$), so at a particular frequency $\omega$ the relation between the fast and slow nonlinearities $\chi^{(3)}_{slow}(\omega) / \chi^{(3)}_{fast}(\omega) \approx 2T_1 / T_2(\omega)$ is conserved.

3. **Carrier motion in the nonparabolic band**

Now, that we have established that the interband nonlinearity has two components – a slow term, associated with excitation of real carriers, and a fast term, associated with virtual carrier excitation – the question is whether the same approach can be applied to the intraband nonlinearity, based the motion of free carriers inside the band, that is responsible for the extraordinary properties of ENZ materials.

To answer this question, the motion of electrons in the isotropic band with dispersion $\mathrm{E}(k)$, where $k$ is a wavevector, is considered when the harmonic field $E_z(t) = E_1 e^{-i\omega_1 t} + E_2 e^{-i\omega_2 t} + c.c.$ is applied along the z-direction (Figure 1c). The equation of motion for a carrier with the wavevector $k_0$ in this band is:

$$\frac{dk_z}{dt} = -\frac{e}{\hbar} E_z - \gamma(k_z - k_{z0}) \tag{10}$$

where $\gamma$ is the momentum scattering rate. The solution for a harmonic input field is $k_z(t) = k_{z0} + \delta k(t)$, where:

$$\delta k(t) = \frac{e}{\hbar(i\omega_1 - \gamma)} E_1 e^{-i\omega_1 t} + \frac{e}{\hbar(i\omega_2 - \gamma)} E_2 e^{-i\omega_2 t} + c.c.. \tag{11}$$

The electron velocity in a the band $\mathrm{E}(k)$ and in the direction of the applied electric field is $v_z = \hbar^{-1}\partial \mathrm{E}/\partial k_z$, and can be expanded into the power series around $k_0$ as:

$$v_z(t) = v_z(k_{z0}) + \frac{\partial v_z}{\partial k_z}\delta k(t) + \frac{1}{2}\frac{\partial^2 v_z}{\partial k_z^2}\delta k^2(t) + \frac{1}{6}\frac{\partial^3 v_z}{\partial k_z^2}\delta k^3(t) + ... \tag{12}$$

Since due to time reversal symmetry $v(k) = -v(-k)$, summation over all the filled states in the band will produce a net zero velocity for the time independent term, we may drop the first term in (12). The first remaining term that is linear in $\delta k(t)$ can then be represented by introducing the (inverse) transport effective mass as:

$$m_t^{-1} = \hbar^{-1} \frac{\partial v_z}{\partial k_z} \qquad (13)$$

so that $\delta v_z = \hbar m_t^{-1} \delta k(t)$. The expression for the transport effective mass is derived in [23]:

$$m_t^{-1} = \hbar^{-1} \left[ \frac{2}{3} \frac{v(k)}{k} + \frac{1}{3} \frac{dv(k)}{dk} \right] = \frac{1}{3\hbar k^2} \frac{d}{dk}\left(k^2 v(k)\right) \qquad (14)$$

This definition is somewhat different from the conventional definition of the effective mass $m_c(k) = \hbar^{-1} dv/dk$ as well as from the alternative definition of the optical effective mass $m_{opt}(k) = \hbar^{-1} v/k$. However, for the case of an ideal parabolic band with a linear velocity-momentum relation, all three definitions result in the same value. For the non-parabolic band, though there is an important difference. Using the definition in (14), when the velocity saturates, i.e. $dv/dk = 0$, the effective mass does not go to infinity. Even when $dv/dk < 0$ $m_t(k)$ stays finite and positive. As we shall see further on, for a typical ENZ material, the difference is not drastic and does not change the conclusion.

Now let us introduce the dispersion of a nonparabolic band derived from a simple two-band **k·P** model. According to Kane [51]:

$$E(k) = \frac{E_{gap}}{2} \sqrt{1 + \frac{2\hbar^2 k^2}{m^* E_{gap}}} = \frac{E_{gap}}{2}\sqrt{1 + (k/k_0)^2} \qquad (15)$$

where $m^*$ is the effective mass at **k**=0, $E_{gap}$ is the bandgap energy and we have introduced the "nonparabolicity" wavevector $k_0 = \sqrt{m^* E_{gap}/2\hbar^2}$ at which the dispersion changes from "parabola-like" to "linear-like" as shown in **Figure 3a**, where the energy is measured relative to the bottom of conduction band, wavevectors are normalized to $k_0$, and the energy is normalized to $E_{gap}/2$.

The velocity, plotted in Figure 3b is:

$$v_z(k) = v_{sat} \frac{(k/k_0)}{\sqrt{1+(k/k_0)^2}} \tag{16}$$

where the saturation velocity is:

$$v_{sat} = \frac{1}{2}\frac{E_{gap}}{\hbar k_0} = \frac{\hbar k_0}{m^*}. \tag{17}$$

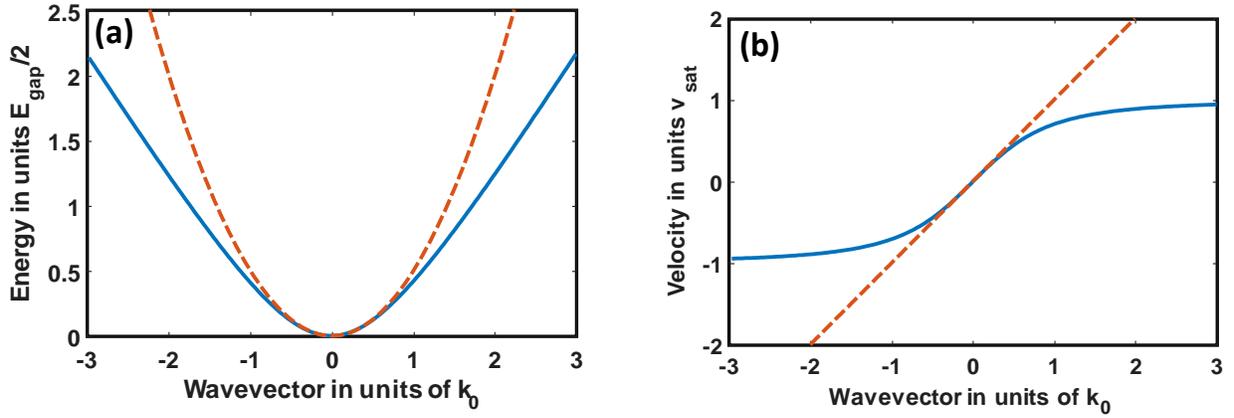

**Figure 3** Dispersion of (a) energy and (b) velocity in the non-parabolic band (solid lines) and its parabolic approximation (dashed lines)

The effective optical mass can be calculated according to (14):

$$m_t^{-1} = \frac{v_{sat}}{\hbar k_0}\frac{2(k/k_0)^2+3}{3\left[1+(k/k_0)^2\right]^{3/2}} = m^{*-1}\frac{2(k/k_0)^2+3}{3\left[1+(k/k_0)^2\right]^{3/2}}; \tag{18}$$

and is shown in **Figure 4a** compared to the other two definitions of the effective mass mentioned above ($m_c(k) = \hbar^{-1} dv/dk$ and $m_{opt}(k) = \hbar^{-1} v/k$). We also compute the first and second derivative of the inverse effective mass as it is required below. The first derivative:

$$\alpha = \frac{dm_t^{-1}}{dk} = \hbar^{-1}\frac{d^2 v_z}{dk^2} = \frac{m^{*-1}}{k_0}\frac{k}{k_0}\frac{2(k/k_0)^2 + 5}{3\left[1+(k/k_0)^2\right]^{5/2}} \tag{19}$$

is plotted in Figure 4b (in units of $m^{*-1}/k_0$), and the second derivative:

$$\beta = \frac{d^2 m_t^{-1}}{dk^2} = \hbar^{-1}\frac{d^3 v_z}{dk^3} = \frac{m^{*-1}}{k_0^2}\frac{4(k/k_0)^4 + 14(k/k_0)^2 - 5}{3\left[1+(k/k_0)^2\right]^{7/2}} \tag{20}$$

is plotted in Figure 4c, normalized to $m^{*-1}/k_0^2$. Note that in their normalized form both coefficients are of the order of unity, and this fact will play an important role further on when we compare the slow nonlinearity that depends on $\alpha$ with the fast one that depends on $\beta$.

From this we can see two possible avenues from which optical nonlinearities can arise, similar to the ideal two-level system described in Section 2. The first avenue is through an applied electric field which effectively polarizes the conduction band electrons in the nonparabolic band (see (12) and Figure 3b). Note that the expansion of carrier velocity, i.e. conductivity current in (12) is conceptually no different from the expansion of polarization (and displacement current) in (6) and (7), and involves no characteristic time. The second avenue is through the absorption of energy which redistributes carriers within the nonparabolic band (see (18) and Fig. 4a). As we will show, the first constitutes the "*fast*" component of the nonlinearity while the second constitutes the "*slow*" effect as its relaxation is governed by the hot carrier lifetime $T_1$.

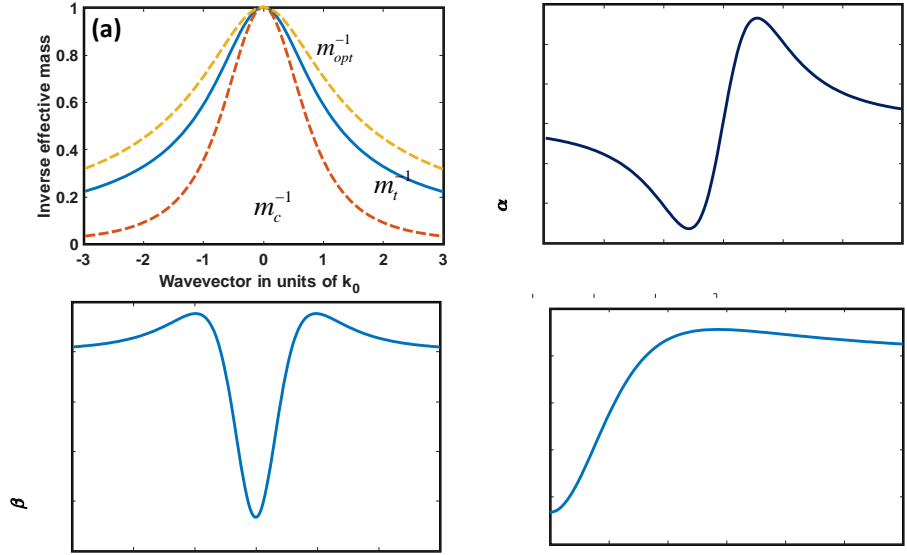

**Figure 4** (a) inverse effective masses $m_t^{-1}$ (solid line), $m_c^{-1}$ and $m_{opt}^{-1}$ (dashed lines) normalized to $m^{*-1}$ (b) normalized first and (c) second derivatives of the inverse effective mass versus $k_0$. (d) mean value of the second derivative versus Fermi wavevector

## 4. Fast nonlinearity in ENZ material

To uncover the fast component of the nonlinearity, we return to (12) and express the electron velocity as a function of the first- and second-derivative of the effective mass:

$$v_z(k,t) = \hbar \left[ m_t^{-1} \delta k(t) + \frac{1}{2} \alpha \delta k^2(t) + \frac{1}{6} \beta \delta k^3(t) \right]. \tag{21}$$

The current density can be then computed as:

$$J(t) = -e \sum_k v_z(k,t) = -e \sum_k^{k_F} \hbar \left[ m_t^{-1} \delta k(t) + \frac{1}{6} \beta \delta k^3(t) \right] \approx -Ne\hbar \left[ \langle m_t^{-1} \rangle \delta k(t) + \frac{1}{6} \langle \beta \rangle \delta k^3(t) \right] \tag{22}$$

where $k_F$ is Fermi wavevector. We note that the summation over $k$ cancels for the second order term as $\alpha(k)$ is an odd function, hence the current has only odd order terms – first order (or linear) $J^{(1)}$ and the third order one $J^{(3)}$. Here $N$ is the carrier density and the averaging is done over the distribution of carriers in the band where:

$$\langle m_t^{-1} \rangle = \frac{3}{k_F^2} \int_0^{k_F} m_t^{-1}(k) k^2 dk = \frac{m^{*-1}}{\left[1+(k_F/k_0)^2\right]^{1/2}};$$

$$\langle \beta \rangle = \frac{3}{k_F^2} \int_0^{k_F} \beta k^2 dk = \frac{m^{*-1}}{k_0^2} \left\{ \frac{4\sinh^{-1}(k_F/k_0)}{(k_F/k_0)^3} - \frac{4(k_F/k_0)^4 + 11(k_F/k_0)^2 + 4}{(k_F/k_0)^2 \left[1+(k_F/k_0)^2\right]^{1/2}} \right\},$$

(23)

The latter is plotted in Figure 4d, and once again, when properly normalized it is of the order of unity. Note that approximation (23) is valid for degenerate doping with a Fermi energy approaching 1 eV, i.e. much larger than the thermal energy of electrons and typical ENZ material operating in visible or near infrared falls into this category.

Substituting (11) into (22) we obtain the linear (first order) current response described by:

$$J^{(1)}(\omega_{1,2}) = \frac{iNe^2 \langle m_t^{-1} \rangle}{\omega_{1,2} + i\gamma} E_{1,2} e^{-i\omega_{1,2} t} + c.c. \tag{24}$$

and the corresponding free carrier part of the susceptibility reads:

$$\chi_{fc}^{(1)}(\omega) = -\frac{Ne^2 \langle m_t^{-1} \rangle}{\varepsilon_0 \omega(\omega + i\gamma)} \tag{25}$$

while the total dielectric constant is $\varepsilon(\omega) = \varepsilon_\infty + \chi_{fc}^{(1)}(\omega)$.

The third-order nonlinear response is:

$$J^{(3)} = -i\frac{Ne^4 \langle \beta \rangle}{6\hbar^2} \left( \frac{E_1 e^{-i\omega_1 t}}{\omega_1 + i\gamma} + \frac{E_2 e^{-i\omega_2 t}}{\omega_2 + i\gamma} + c.c. \right)^3 \tag{26}$$

Selecting the FWM term we get:

$$J_{FWM}^{(3)} \approx -i \frac{Ne^4 \langle \beta \rangle}{2\hbar^2 (\bar{\omega}^2 + \gamma^2)^{3/2}} \left( E_1^2 E_2 e^{-i(2\omega_1 - \omega_2)t} + E_2^2 E_1 e^{-i(2\omega_2 - \omega_1)t} \right) e^{i\Phi} + c.c. \quad (27)$$

and the "*fast*" FWM susceptibility:

$$\chi_{fast}^{(3)}(\omega_3 = \omega_1 - \omega_2 + \omega_1) = \frac{Ne^4 \langle \beta \rangle}{2\varepsilon_0 \hbar^2 (\bar{\omega}^2 + \gamma^2)^{3/2}} \approx \chi_{fc}^{(1)}(\omega) \frac{e^2 f_\beta}{2\hbar^2 k_0^2 (\bar{\omega}^2 + \gamma^2)} \quad (28)$$

where

$$f_\beta(k_F) = k_0^2 \langle \beta \rangle \langle m_t \rangle \quad (29),$$

is plotted in **Figure 5**, and is once again on the order of unity. Obviously, equation (28) describes the fast nonlinearity as its magnitude does not depend on the beat frequency $\Delta\omega = \omega_1 - \omega_2$. Additionally, it can been seen from (26) that the third harmonic susceptibility will have a similar magnitude, which would not be the case for the slow component.

To evaluate the reasonability of the approach, we may consider ITO with a carrier density $N = 1.1 \times 10^{21} cm^{-3}$ i.e. $k_F = (3\pi^2 N)^{1/3} = 3.2 nm^{-1}$. The value of $k_0$ can be obtained by fitting as roughly $2 nm^{-1}$ which makes $k_F / k_0 \approx 1.6$ and $f_\beta \approx 0.5$. Furthermore, since we are concerned with the response near the ENZ condition, $\chi_{fc}^{(1)}(\omega) \sim -\varepsilon_\infty \approx -4$ and so for 1 eV excitation we obtain:

$$\chi_{fast}^{(3)}(\omega_3 = \omega_1 - \omega_2 + \omega_1) \approx \varepsilon_\infty \frac{e^2}{2\hbar^2 \omega^2 k_0^2} \langle f_\beta \rangle \approx 2.5 \times 10^{-19} m^2 / V^2 \quad (30)$$

This instant nonlinearity due to free electrons has roughly the same order of magnitude as the fast non-resonant nonlinearity due to bound electrons as estimated in (9). Comparing (30) with (9) one can notice that $\varepsilon_\infty$ for ENZ material is comparable to $\chi^{(1)}$ of traditional material, therefore it appears that $z_{21}$ and $k_0$ are related as $z_{12}k_0 \approx 1$. We shall return with an explanation of this remarkable fact in Section 6.

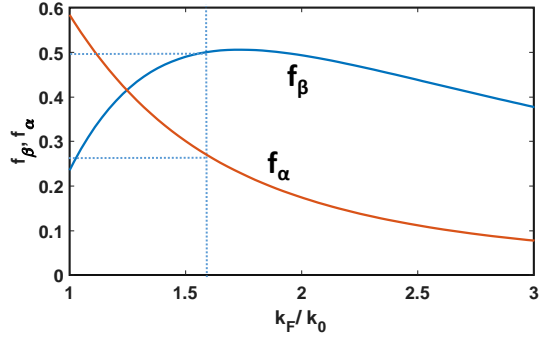

**Figure 6** The coefficients $f_\beta$ and fast nonlinearity and $f_\alpha$ for slow nonlinearity as functions of the Fermi level position.

## 5. Slow (thermal) nonlinearity in ENZ material

Next, we move to an estimate of the "*slow*" nonlinearity. To do so, we must calculate the amount of absorbed power per unit volume, found from (24) as:

$$\delta P(t) = J(t)E(t) = \frac{2Ne^2\gamma\langle m_t^{-1}\rangle}{\bar{\omega}^2 + \gamma^2}\left[E_1^2 + E_2^2 + 2E_1E_2\cos[(\omega_1 - \omega_2)t]\right], \tag{31}$$

where the oscillations at frequencies of the order $2\omega$ are neglected. The absorbed power is transferred to the hot electrons excited from below to above Fermi level as shown in Figure 1c, left panel. The excess energy density of hot carriers can be found from:

$$\frac{dU_{hot}(t)}{dt} = \delta P(t) - \frac{U_{hot}}{\tau_{el}} \tag{32}$$

where $\tau_{el}$ is the energy relaxation rate between hot carriers and lattice. We are interested in the energy density oscillations at beat frequency $\Delta\omega = \omega_1 - \omega_2$:

$$U_{hot}(t) = \frac{\tau_{el}}{1-i\Delta\omega\tau_{el}} \frac{2Ne^2\gamma\langle m_t^{-1}\rangle}{\bar{\omega}^2+\gamma^2} E_1 E_2 e^{-i(\omega_1-\omega_2)t} + c.c. \tag{33}$$

This energy is not shared equally by all the conduction electrons as the hot electrons include both the "primary", generated when photon is absorbed, and the "secondary" carriers, generated via fast (100fs or less) electron-electron collisions. Thus the number of excited carriers is not fixed. Usually it is assumed that the electrons thermalize with a certain electron temperature $T_e$ which needs to be evaluated. However, even though electron-electron collissions are fast, it may take longer than 100 fs to establish thermal equilibrium. Therefore, it is preferable to evaluate the nonlinearity without making any assumption of thermal equilibrium.

Let us say the fraction of the "hot" carriers promoted from below the Fermi levels is $f_{hot}$ and their density is $f_{hot}N$. Then the energy of the average hot carrier oscillates as $\delta E(t) = U(t)/f_{hot}N$, and the average time-dependent change of the wavevector is:

$$\delta k(t) = \delta E(t)/\hbar v_F, \tag{34}$$

which causes the change of the inverse effective mass of the hot carriers:

$$\delta m_t^{-1}(t) = \frac{dm_t^{-1}}{dk}\delta k(t) = \alpha_F \frac{U_{hot}(t)}{f_{hot}N\hbar v_F}, \tag{35}$$

where $\alpha_F = |\alpha(k_F)|$, and describes first derivative of inverse transport effective mass af Fermi level. The change in susceptibility, and hence dielectric constant, can then be estimated as:

$$\delta\varepsilon(t) = \delta\chi^{(1)}(t) \approx f_{hot}\frac{\delta m_t^{-1}}{\langle m_t^{-1}\rangle}\chi^{(1)} = \alpha_F \frac{U_{hot}(t)}{\langle m_t^{-1}\rangle N\hbar v_F}\chi^{(1)}, \tag{36}$$

where the transport mass is averaged over the electron distribution. Note that, as expected, the change in the susceptibility does not depend on $f_{hot}$. Hence, as long as the electron temperature is not excessively high, $T_e \ll E_F / k_B \approx 15,000K$, it is not required to compute the exact distribution of the carriers that are excited to obtain a decent estimate of nonlinearity, and the calculation of the electron temperature is, therefore, unnecessary. Substituting (33) we obtain the beat frequency oscillations of the dielectric constant:

$$\delta\varepsilon(\omega_1 - \omega_2) \approx \alpha_F \frac{1}{N\hbar v_F} \frac{\gamma \tau_{el}}{1 - i\Delta\omega\tau_{el}} \frac{2Ne^2}{(\bar{\omega}^2 + \gamma^2)} \chi^{(1)} E_1 E_2 e^{-i(\omega_1 - \omega_2)t} + c.c. \tag{37}$$

When the electric field $E_1 e^{-i\omega_1 t}$ scatters from these oscillations, the nonlinear polarization arises, with a "*slow*" FWM susceptibility given by:

$$\chi^{(3)}_{slow}(\omega_3 = \omega_1 - \omega_2 + \omega_1) = \frac{\gamma \tau_{el}}{1 - i\Delta\omega\tau_{el}} \frac{2\alpha_F}{\hbar v_F} \frac{e^2}{\bar{\omega}^2 + \gamma^2} \chi^{(1)} \tag{38}$$

From (16) and (19) we get

$$\frac{\alpha_F}{v_F} = \frac{m^{*-1}}{v_{sat} k_0} f_\alpha(k_F) = \frac{1}{\hbar k_0^2} f_\alpha(k_F), \tag{39}$$

where

$$f_\alpha(k_F) = \frac{2(k_F / k_0)^2 + 5}{3\left[1 + (k_F / k_0)^2\right]^2} \tag{40}$$

and is plotted in **Figure 6**, and once again is on the order of unity. Substituting (39) into (38) we obtain the final expression for the slow nonlinearity:

$$\chi^{(3)}_{slow}(\omega_3 = \omega_1 - \omega_2 + \omega_1) = \frac{2\tau_{el}/\tau_s}{1-i\Delta\omega\tau_{el}} \chi^{(1)} \frac{e^2 f_\alpha}{\hbar^2 k_0^2 (\bar{\omega}^2 + \gamma^2)}, \qquad (41)$$

where $\tau_s = \gamma^{-1}$ is the momentum scattering time. Momentum scattering typically occurs due to scattering on impurities and phonons and can be estimated from the mobility measurements. For ITO, AZO and many other transparent oxides it is on the scale of 10 fs or less. The electron-lattice relaxation time $\tau_{el}$ is also determined by the electron-phonon scattering, but is typically much longer than $\tau_s$ and is on the scale of 100's of femtoseconds. The reason for it is three-fold. First of all, momentum scattering occurs independently of whether the scattering event involves the absorption or emission of a phonon, but the energy transfer from hot carriers to lattice is determined by the net emission of phonons, i.e. the difference of phonon emission and absorption rates. Second, the phonon energy is usually less than thermal energy of hot carriers $k_B T_e$, so it takes quite a few phonon emission events per carrier to cool them down. Finally, the impurity and surface scattering contributing to momentum relaxation are elastic processes and do not provide a channel for energy transfer to the lattice.

From **Figure 6** one can note that for the example of ITO with $k_F/k_0 = 1.6$, $f_\alpha \approx f_\beta/2$, and for other ENZ materials there is not much difference. Therefore one can write the relation between the slow (41) and fast (28) nonlinearities as:

$$\chi^{(3)}_{slow} \approx \frac{2\tau_{el}/\tau_s}{1-i\Delta\omega\tau_{el}} \chi^{(3)}_{fast}, \qquad (42)$$

*which is quite similar to the relation between the slow (real) and fast (virtual) interband nonlinearities in the more conventional materials discussed in Section 2, see (8).* It follows that $\chi^{(3)}_{slow}$ is about two order of magnitude higher than $\chi^{(3)}_{fast}$, up to $10^{-17} m^2/V^2$. Clearly, the momentum

scattering time $\tau_s$ plays the role of the dephasing time $T_2$ and the electron-lattice energy relaxation time $\tau_{el}$ plays the role of the recombination time $T_1$. But the magnitudes of these times are quite different in ENZ materials. Typical values of the recombination times in dielectrics and semiconductors are on the nanosecond scale, while $\tau_{el}$ in TCOs is less than a picosecond, hence even the "slow" nonlinearity in TCO can be considered "ultrafast" as the term "ultrafast" is defined today (sub-picosecond, or THz). It is quite conceivable that as ultrafast science progresses, the definition of "ultrafast" may shift into femtosecond domain. But, of course, in TCOs the nonlinearity enhancement is weaker than in the nonlinearity based on saturation of absorption – the usual gain-bandwidth compromise.

**Discussion**

Through this discussion it is clear that there is a strong connection between the nonlinearities in traditional materials and ENZ materials. Let us now consider how the nonlinearity can be tailored for a given application.

The fast nonlinearity does not radically change whether it be from one material to another or if one considers traditional nonlinearities due to bound carriers, or nonlinearities due to free carriers in ENZ materials. We can understand this by considering the terms that play into the strength of the fast nonlinearity – for odd order traditional nonlinearities only the matrix element of the dipole transition between valence and conduction bands $\mu_{cv} = ez_{cv}$, the density, and the detuning from resonance play a role. Among them, the dipole $z_{cv}$ is proportional to the bond length and the density is inversely proportional to the its cube, but in all nonlinear materials the bond length remains roughly constant varying between 1.6 and 2.1 Angstrom [52]. Similarly, the effective detuning $\bar{\omega}_{21} - \omega$ is commensurate with bandgap, which is again roughly the same for the materials with the same transparency bandwidth. Together, this explains why the magnitude of nonlinearities in dielectrics are all similar and only show some increase with an increase in wavelength as one can use materials with narrower bandgap.

For the nonlinearities in ENZ materials, the magnitude depends upon the density of free carriers and the band nonparabolicity. Although the free carrier density is not fixed, if one is to approach the ENZ condition in the telecommunications spectrum (as is the case for most TCOs) the density must be high enough to cancel the positive dielectric constant which forces the majority of materials to maintain a comparable carrier density $\sim 10^{21}$ $cm^{-3}$. What is important though is that linear susceptibility of the ENZ material due to free carriers is comparable to the linear susceptibility of traditional nonlinear materials. Additionally, the nonlinearity hinges on the non-parabolicity of the band, as expressed by the value of $k_0^{-1}$ in (15). However, the strength of the interband nonlinear and the band nonparabolicity are connected by the same oscillator sum rule [53]:

$$\frac{2}{m_0} \sum_{k \neq n} \frac{P_{kn}^2}{E_k - E_n} + \frac{m_0}{m_n^*} = 1, \qquad (43)$$

where the first term contains the strength of all the interband transitions originating in the band n and $m_n^*$ is the effective mass of that band. For the two band model, using definition of $k_0$ this expression can be simplified to:

$$\frac{m_0 E_{gap}}{2\hbar^2 k_0^2} = 1 + \frac{2P_{cv}^2}{m_0 E_{gap}} \qquad (44)$$

where $P_{cv}$ is the matrix element of the momentum for the valence-to-conduction band transition given by $P_{cv} = m_0 z_{cv} E_{gap} / \hbar$ according to Kane's $\mathbf{k} \cdot \mathbf{P}$ [51] theory. Combining these two expressions we can see that $k_0^{-1} \sim z_{cv}$. As a result, fast nonlinearities in traditional and ENZ materials are both determined by the strength of the interband transition, and thus it follows that their magnitudes are quite similar.

The main difference between fast intraband (free carriers), and interband (bound carriers) nonlinearities is that the loss in the intraband nonlinearities is unavoidable simply because for any photon frequency there always exist plenty of filled initial and empty final states. In the interband nonlinearities, as long as one operates far enough from the absorption edge the loss is minimized. That explains why for the femtosecond nonlinear processes, such as super-continuum generation and optical frequency combs, or for the harmonic generation, the low loss transparent materials, such as $SiO_2$, SiN, diamond, and $LiNbO_3$ remain dominant and it would be a trall order for lossy materials like ENZ will replace them.

As for slow nonlinearities, they rely on absorption which provides a broad range of tunability in the nonlinear response for both the traditional and ENZ nonlinearities, with ENZ materials, as explained below, having distinct advantages. To identify the optimal material properties, we approximate the propagation (absorption) length as $L_a \sim c\chi^{(1)} / \gamma$ which can be used to introduce a figure of merit (FOM) - the product of maximum nonlinearity, propagation length, and the signal bandwidth B – to evaluate and optimize mateirals. For ENZ materials the FOM becomes:

$$FOM = \chi_{slow}^{(3)} L_a B \approx \frac{n_g e^2 c f_\alpha}{2\pi \hbar^2 k_0^2 \bar{\omega}^2}, \; B = 1/2\pi\tau_{el} \;, \tag{45}$$

and for the traditional nonlinearity:

$$FOM = \chi^{(3)}_{slow} L_a B \approx \frac{n_g e^2 z_{21}^2 c\bar{\omega}}{2\pi \hbar^2 (\omega_{21} - \bar{\omega})^3}, \quad B = 1/2\pi T_1, \tag{46}$$

where an additional enhancement by the group index $n_g$ has been included due to enhanced interaction time between light and matter [32, 54]. For ENZ materials this can be a factor of a few in thin film form, while for most traditional nonlinearities, a large group index can be attained only by utilizing specially fabricated structures such as photonic crystals.

As one can see, the FOM is rather well defined and constant for a given wavelength, therefore a prudent way to optimization leads to making the bandwidth just sufficient for a given task. In this sense, the ENZ response time, $100 fs < \tau_{el} < 1 ps$, is almost ideal to allow the nonlinear effects to accumulate over the time $\tau_{el}$ while remaining fast enough for many applications in telecommunications such as all-optical switching, adiabatic frequency shifting, and FWM at a speed of a few THz.

For the interband transitions, the recombination time $T_1$ is about 3 orders of magnitude longer for direct bandgap materials (and even longer for indirect bandgap), and that is way too long for all optical processing, but may be sufficiently fast for spatial light modulators and or all optical routers. Of course, the speed of the interband nonlinearity can be enhanced if the recombination time is quenched in structures full of defects, for instance low temperature growth GaAs [55], but that introduces significant background absorption, and the only application of these materials is as saturable absorbers, which of course may be extremely useful for some applications, like passive mode-locking [56], and for photoconductive switching for the generation of THz waves.

Aside from having just about the ideal response time, an important advantage of the ENZ "slow" nonlinearity is that it is fairly broadband (in the sense of *optical bandwidth B$_{opt}$*) since the absorption follows the smooth Drude dispersion $\sim \omega^{-2}$. That is dramatically different from the traditional slow nonlinearity whose magnitude may change by orders of magnitude over a few tens of nanometers. Of course, since the absorption length also changes by the equal amount, the FOM is broadband, but that means that the length of the device should be changed when wavelength changes, which makes it impractical. Therefore, as mentioned above slow interband nonlinearities are used almost exclusively as saturable absorbers [56], an important, but a relatively narrow application niche.

Before concluding, one can also make an interesting comparison with intersubband nonlinearities in semiconductor quantum wells, investigated at length in 1990's [57, 58] but given a new life more recently in combination with metasurfaces [59, 60]. On one hand these are intraband nonlinearities (like ENZ), but on the other hand the carriers in them are confined and the transitions are between discrete levels. However, the intersubband relaxation time $T_1$ and intrasubband scattering time $T_2$ in square QW's are roughly of the same order of magnitude (100's of fs), while the lineshape is perfectly Lorentzian. As a result, in simple structures there is no huge difference between the "slow" and "fast" nonlinearities in terms of magnitude, although by means of band engineering one can increase the intersubband relaxation time and enhance the slow nonlinearity [61, 62], obviously at the price of making it slower, but still sufficiently fast for THz applications. In that respect intersubband nonlinearities and nonlinearities in ENZ are very similar (although intersubband nonlinearities are currently limited to mid-IR by the finite values of band offsets in semiconductors). The main difference lies in the fact that intersubband transitions are relatively narrow and thus one can provide additional resonant enhancement, but obviously at the expense of optical bandwidth, while ENZ nonlinearities are inherently broadband. Therefore, rather than competing, ENZ and intersubband nonlinearities complement each other depending on what is more important: sheer magnitude of the effect or optical bandwidth, and what spectral range is being considered.

## 6. Conclusions

In this study we have placed the nonlinearity in ENZ materials based on conductive oxides into the conventional framework of nonlinear susceptibilities. While the mechanism of nonlinearity in ENZ materials (band nonparabolicity and carrier heating) is seemingly different than the mechanism in more conventional nonlinear materials, ENZ nonlinearities can still be expressed in terms of nonlinear susceptibilities. Just as in conventional materials, there are two components to the effect – a "fast", or "instant", and a relatively "slow" one. Despite apparently different mechanisms, the magnitudes of "fast" nonlinearity at a given wavelength for the ENZ and conventional nonlinearity are roughly of the same magnitude. A deeper look reveals that in the end, intraband ENZ and conventional interband mechanisms rely on the same dipole matrix element of the interband transition to achieve their fast nonlinearity, hence their similar magnitudes (for a given wavelength) should not come as a surprise.

Another important conclusion is that, for both the nonlinearity in ENZ and conventional nonlinearities, the relative strength of slow and fast components is determined by the ratio of two characteristic times, $T_1/T_2$. $T_1$ is the energy relaxation time that determines the speed of the "slow" nonlinearity while $T_2$ is momentum relaxation (or dephasing) time, which also determines the absorption strength. *As a result, the figure-of-merit, defined as the product of nonlinear susceptibility, signal bandwidth and absorption length, is roughly the same for any nonlinear mechanism for a given wavelength. However, ENZ does have three important advantages.*

First of all, the energy relaxation time in ENZ being a few hundred femtoseconds, happens to be just right for many important applications, which means that the relatively "slow" ENZ nonlinearity is still ultrafast on the absolute scale. In other words, for THz scale processes the nonlinearity in ENZ is as high as it can be (for a given wavelength). Second, the "slow" nonlinearity in ENZ materials is very broadband due to the non-resonant character of absorption. And finally, both the fast and slow nonlinearities in ENZ are enhanced by low group velocity achieved without any required nanofabrication. As a result, many experimental results achieved in the most recent past have made use of these unique benefits to demonstration exceptional nonlinear interactions in ENZ thin films including adiabatic frequency conversion [32], negative refraction [29, 30] and phase conjugation, all-optical switching [26, 28], bi-color switching [63], and more [64, 65]

Subsequently de-mystified, ENZ materials may not be a magic solution for all tasks nonlinear optics is expected to perform, but they do have an important niche to occupy – the niche in which one is looking for a temporal response on the scale of a few hundreds of femtoseconds and compact size. Combined with the fact that many ENZ materials achieve these properties in the near infrared spectral range using well-established CMOS-compatible materials bodes well for the continued exploration for THz switching and routing in telecommunications as well as compact wave mixing and harmonic generation in both integrated and free-space platforms. As a result, in contrast to the numerous promising platforms explored in the past which have come and gone, we expect ENZ materials to be a mainstay in the nonlinear community for the foreseeable future and continue driving advancements in key application spaces of nonlinear optics.

**Acknowledgments**

J.K. ackmowleges the support  from DARPA NLM program HR00111820063.  N.K. acknowledges the support from the Air Force Office of Scientific Research (FA9550-1-18-0151), the National Science Foundation (grant 1808928) and the Virginia Commonwealth Cyber-security Initiative (xxxxx). M.C. acknowledges the support from the UK Research and Innovation - Engineering and Physical Sciences Research Council (EP/S001573/1) and the Royal Society (RGS\R1\201365).

**Conflict of interests**

The authors declare no competing financial interest.